\documentclass[article]{aa}
\usepackage{graphicx}
\usepackage[varg]{txfonts}
\begin{document}

\title{Deriving physical parameters of unresolved star clusters}
\subtitle{III. Application to M31 PHAT clusters}
\author{P. de Meulenaer\inst{1,2} \and D. Narbutis\inst{2} \and T. Mineikis\inst{1,2} \and V. Vansevi\v{c}ius\inst{1,2}}

\institute{Vilnius University Observatory, \v{C}iurlionio 29, Vilnius LT-03100, Lithuania\\
\email{philippe.demeulenaer@ff.stud.vu.lt} \and
Center for Physical Sciences and Technology, Savanori\c{u} 231, Vilnius LT-02300, Lithuania}

\date{Received 7 October 2014 / Accepted 3 December 2014}

\abstract
{This study is the third of a series that investigates the degeneracy and stochasticity problems present in the determination of physical parameters such as age, mass, extinction, and metallicity of partially resolved or unresolved star cluster populations situated in external galaxies when using broad-band photometry.}
{This work tests the derivation of parameters of artificial star clusters using models with fixed and free metallicity for the WFC3+ACS photometric system. Then the method is applied to derive parameters of a sample of 203 star clusters in the Andromeda galaxy observed with the HST.}
{Following Papers I \& II, the star cluster parameters are derived using a large grid of stochastic models that are compared to the observed cluster broad-band integrated WFC3+ACS magnitudes.}
{We derive the age, mass, and extinction of the sample of M31 star clusters with one fixed metallicity in agreement with previous studies. Using artificial tests we demonstrate the ability of the WFC3+ACS photometric system to derive the metallicity of star clusters. We show that the metallicity derived using photometry of 36 massive M31 star clusters is in a good agreement with the metallicity previously derived using spectroscopy taken from literature.}
{}


\maketitle

\section{Introduction}

Several recent studies \citep[e.g.,][ hereafter Paper I]{Fouesneau2010,Popescu2010,Anders2013,de_Meulenaer2013} were aimed at the determination of partially resolved or unresolved star cluster parameters based on broad-band photometry and using more elaborated models than the traditional simple stellar population (SSP) models, strongly biased by the presence of stochasticity problem in star clusters. The stochastic presence of massive stars in a star cluster causes its integrated colors to be dispersed, making SSP model predictions biased, especially in case of low-mass star clusters. Using a grid of star cluster models in which stellar masses are randomly sampled to take stochasticity into account, \citet[][hereafter Paper II]{de_Meulenaer2014} showed how it is possible to determine the parameters when the metallicity of clusters is unknown, and also quantified the accuracy of the metallicity derivation itself.

In this study we apply these stochastic models to real clusters from the Andromeda galaxy. Using the Wide Field Camera 3 (WFC3) and the Advanced Camera for Surveys (ACS) on board the Hubble Space Telescope (HST), the Panchromatic Hubble Andromeda Treasury (PHAT) team \cite[see, e.g.,][]{Dalcanton2012,Beerman2012,Weisz2013} performed a photometric survey of 1/3 of the M31 galaxy, providing a large catalog of new clusters. \cite{Fouesneau2014} derived the age, mass, and extinction of the PHAT ``year 1'' cluster catalog using the aperture photometry data provided by \cite{Johnson2012}.  

The objective followed here is to study a sample from the PHAT cluster catalog allowing variable metallicity in the model grid to explore the metallicity effects by comparing the results to the case with the fixed metallicity. We also aimed to derive the metallicity of objects with reasonably good photometric accuracy.

	The structure of the paper is the following: Section \ref{sec:Artificial_tests} presents the method of star cluster parameter determination and tests its accuracy using the WFC3+ACS photometric system. Section \ref{sec:PHAT_clusters} presents the application of the method on the sample of the PHAT star clusters. Conclusions are presented in Section \ref{sec:conclusions}.

\section{The method of parameter derivation}
\label{sec:Artificial_tests}
\subsection{Presentation of the method}

\begin{figure*}
\centering
\includegraphics[scale=0.55]{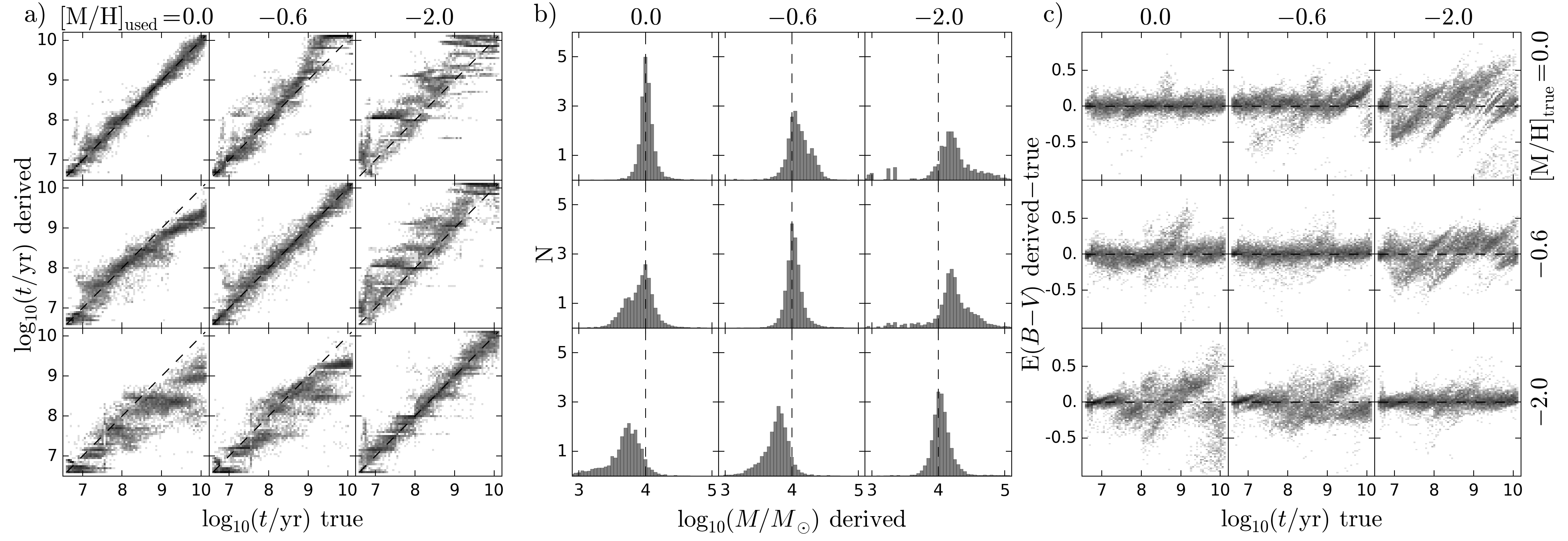}
\caption{\small Age (panel \textbf{a}), mass (panel \textbf{b}), and extinction (panel \textbf{c}) derived for a sample of $10\,000$ artificial star clusters with true mass $\log_{10}(M/M_{\odot}) = 4$ using WFC3+ACS photometry with $\sigma$ = 0.05 mag for the Gaussian photometric errors for each filter. The true metallicity of the artificial clusters is specified on the right and the metallicity of the grid used to classify the clusters is specified on the top.}
\label{fig:1Z_vs_1Z_results_WFC3}
\end{figure*}

As presented in Papers I \& II, the determination of the physical parameters (age, mass, extinction, and metallicity\footnote{We refer to extinction and metallicity as E($B-V$) and $\mathrm{[M/H]}$ hereafter.}) of a given observed star cluster is based on a comparison of its integrated broad-band photometry to a grid of star cluster models. 

The first task is thus to build a 4--dimensional grid of models for every value of the four physical parameters. 
To take the stochasticity problem into account, we computed the grid of discrete cluster models by randomly sampling the stellar mass according to the initial mass function \citep[IMF,][]{Kroupa2001} following the method described in \cite{Deveikis2008} \citep[see also][]{Santos1997,Cervino2002}. The stellar luminosities are extracted from isochrones of the selected age and metallicity of the model. We used the PADOVA isochrones\footnote{PADOVA isochrones from ``CMD 2.6'': http://stev.oapd.inaf.it/cmd} from \cite{Marigo2008} with the addition of TP-AGB phase from \cite{Girardi2010}. Although more recent models with revised solar metallicity are available \citep[PARSEC;][]{Bressan2012, Chen2014}, we did not select them as the TP-AGB phase is not yet included. The grid was built according to the following nodes: from $\log_{10}(t/\mathrm{yr})=6.6$ to 10.1 in steps of 0.05, from $\log_{10}(M/M_{\odot})=2$ to 7 in steps of 0.05, and for 13 metallicities: from $\mathrm{[M/H]}$ = $+0.2$ to $-2.2$ in steps of 0.2. This results in a grid of 71 values of age, 101 values of mass, with $1\,000$ models per node, hence $\sim$$7\times10^{6}$ models for each metallicity. To limit the number of models that need to be stored in computer memory, the extinction was computed when the observed cluster was compared with the grid of models. It ranges from $E(B-V)=0$ to 1 in steps of 0.01, therefore 101 values for the extinction. We used the Milky Way standard extinction law from \cite{Cardelli1989}. 
 	
In a similar way to \cite{Fouesneau2010}, \cite{Fouesneau2014}, and Paper II, we evaluated the likelihood of each node of the grid to represent the observed magnitudes of the cluster. Within each node, we first computed the likelihood of each of the $1\,000$ star cluster models by
\begin{equation}
L_{\mathrm{model}} = \prod_{f=1}^{F} \frac{1}{ \sqrt{2 \pi}\, \sigma_{f}} \exp \left[ - \frac{\left(\mathrm{mag}_{f,\mathrm{obs}}-\mathrm{mag}_{f,\mathrm{model}}\right)^{2}}{2\,\sigma_{f}^{2}} \right]\,,
\end{equation}
where $f$ stands for one particular filter, $\mathrm{mag}_{f}$ for the observed and model magnitudes in that filter, and $F$ for the total number of filters. For example, $F=6$ for the WFC3+ACS photometric system we use in this study. Then the likelihood of the node of age $t$, mass $M$, extinction $E(B-V)$, and metallicity $\mathrm{[M/H]}$ is the sum of the likelihoods of its models,
\begin{equation}
L_{\mathrm{node}}\left(t,M,E(B-V),\mathrm{[M/H]}\right) = \sum_{n=1}^{N} L_{\mathrm{model},\,n}\,,
\label{eq:node_likelihood}
\end{equation}
where $N=1\,000$, the total number of models contained in the node. The procedure is repeated for each node of the 4--dimensional grid, and the observed star cluster is then classified with the parameters of the node, which maximizes the quantity $L_{\mathrm{node}}$.

\subsection{1--metallicity vs 1--metallicity: exploration of the metallicity effects} 
   
\begin{figure*}
\centering
\includegraphics[scale=0.64]{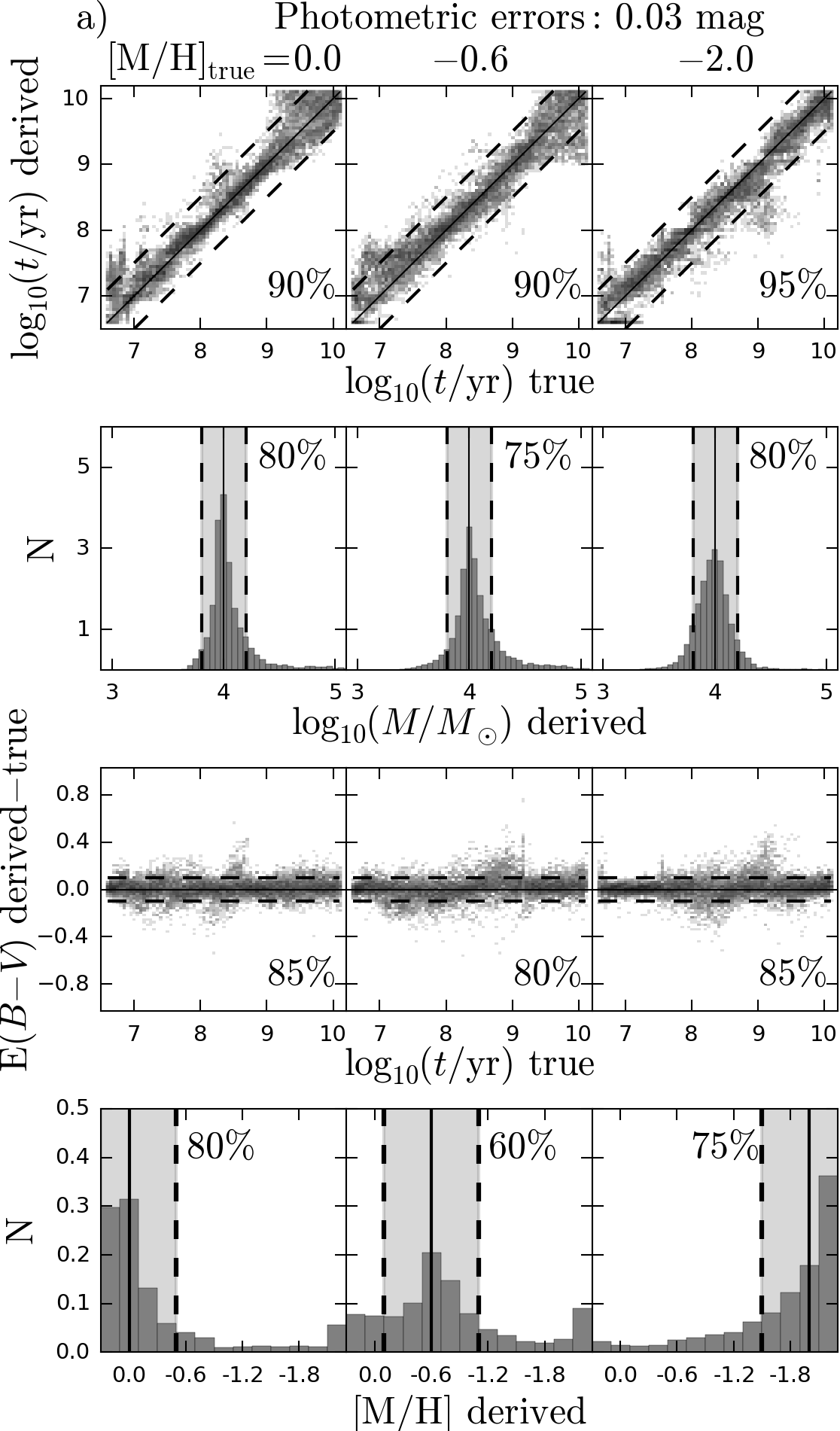}
\includegraphics[scale=0.64]{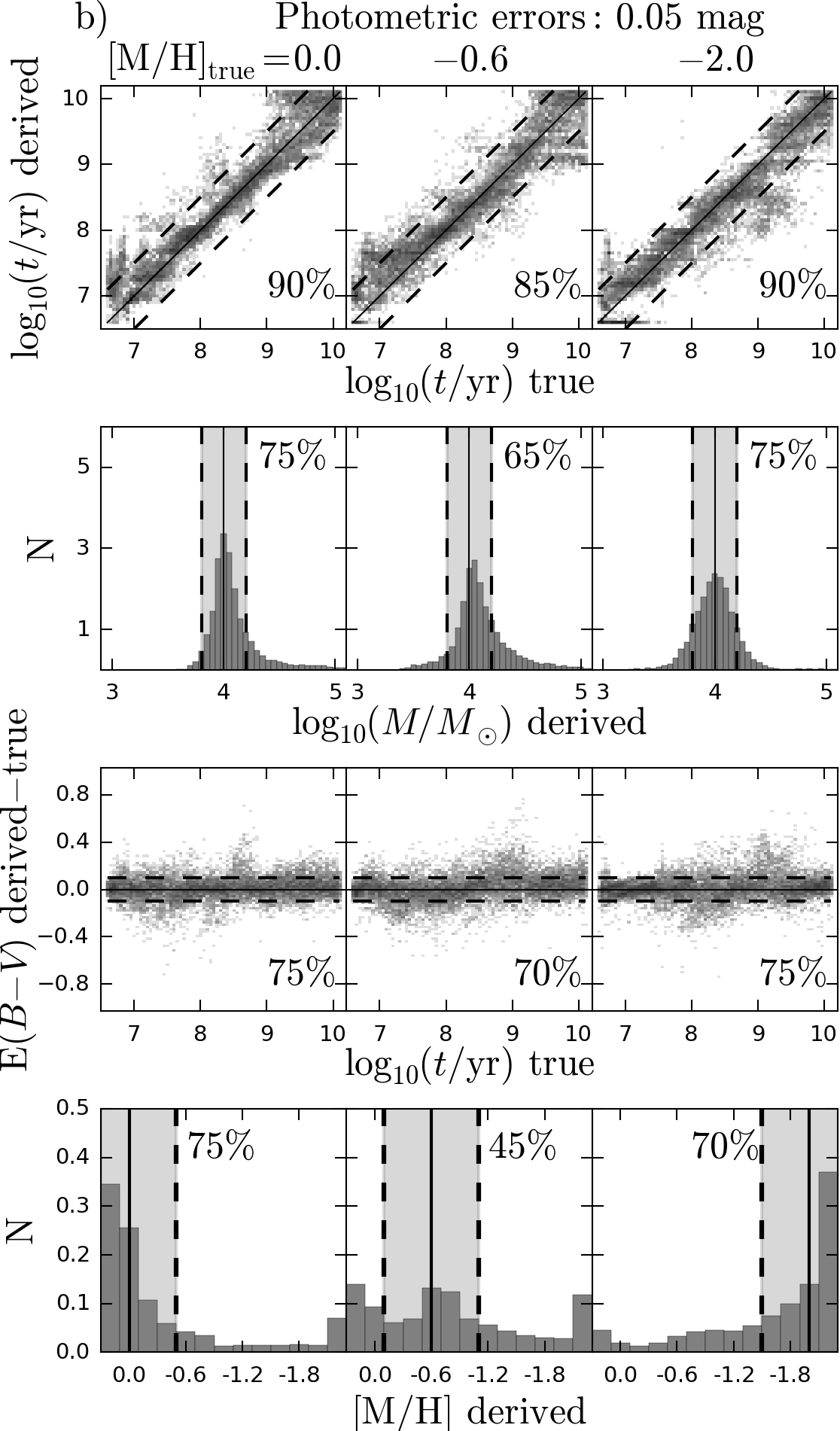}
\includegraphics[scale=0.64]{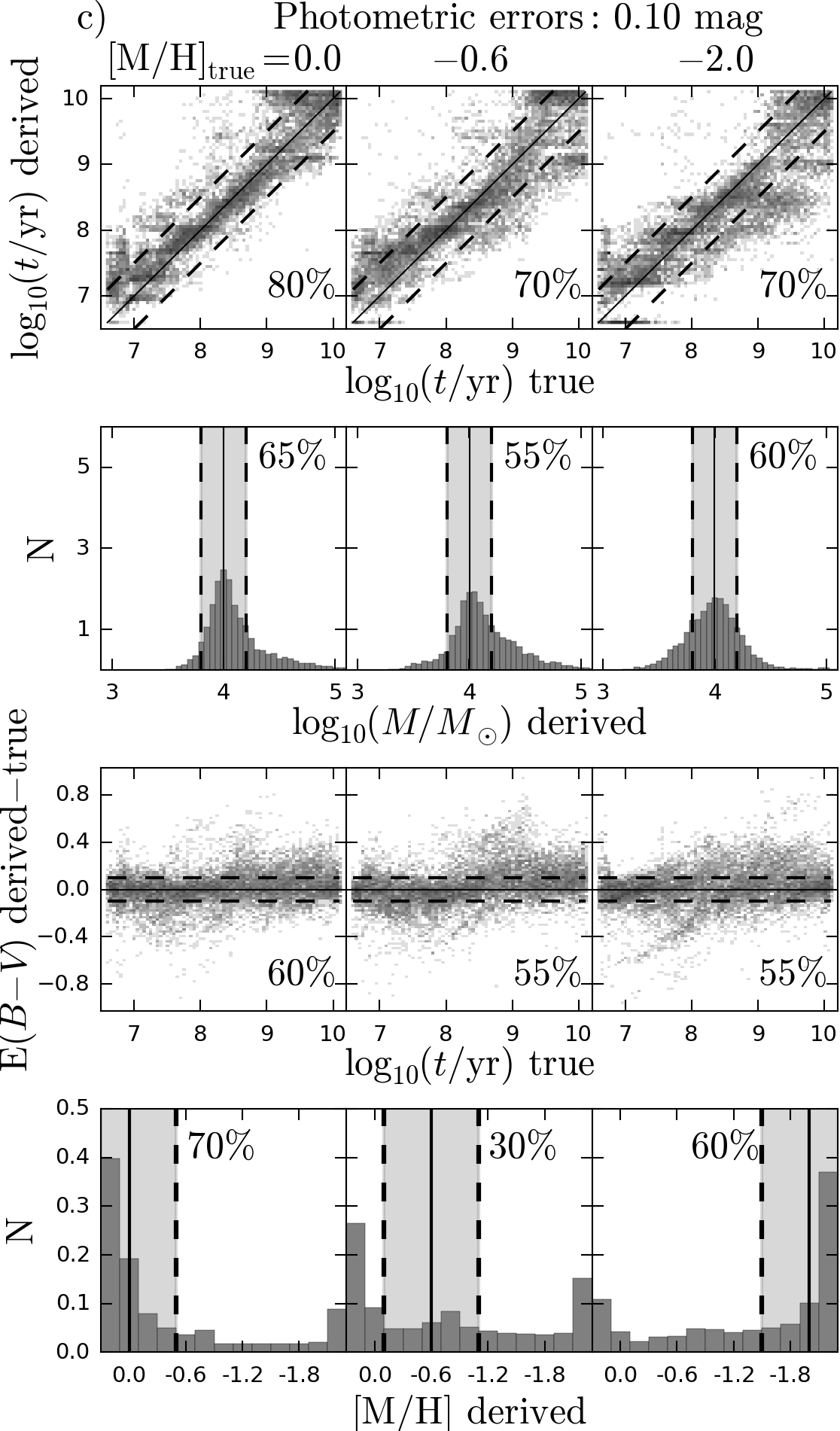}
\caption{\small Age, mass, extinction, and metallicity derived for a sample of $10\,000$ artificial star clusters of true mass $\log_{10}(M/M_{\odot}) = 4$ , with $\sigma$ = 0.03 mag of Gaussian photometric errors (block of panels \textbf{a}), $\sigma$ = 0.05 mag (block \textbf{b}), and $\sigma$ = 0.1 mag (block \textbf{c}) using WFC3+ACS broad-band photometry. The true metallicity of the artificial clusters is indicated on the top of each column of panels. In each panel, the numbers indicate the percentage of the 10\,000 clusters with derived parameter in the region between the dashed lines, which is centered on the true parameter value.}
\label{fig:1Z_vs_3Z_results_WFC3}
\end{figure*}

We apply the method of star cluster parameter derivation using the WFC3+ACS photometric system: UVIS1/F275W, UVIS1/F336W, ACS/F475W, ACS/F814W, IR/F110W, and IR/F160W to characterize the ability of this photometric system to derive the physical parameters of clusters. 

We generated three samples of 10\,000 artificial star clusters with random age in the range $\log_{10}(t/\mathrm{yr})=[6.6,10.2]$, mass fixed to $\log_{10}(M/M_{\odot})$ = 4, with random reddening in the range E($B-V$) = [0, 1]. Each of the three samples has a fixed metallicity, $\mathrm{[M/H]}$ = 0, $-0.6$, or $-2$. Gaussian photometric errors of $\sigma$ = 0.05 mag were added to the magnitudes of the artificial clusters for each filter. 

In this test, we derived the age, mass, and extinction of clusters using model grids of fixed metallicity ($\mathrm{[M/H]}$ = 0, $-0.6$, or $-2$), and the results are displayed in Fig.\,\ref{fig:1Z_vs_1Z_results_WFC3}. In panel a), we see that using a metal poor model grid to analyze a set of metal rich clusters can result in overestimated derived ages by about a factor 10. This is the well-known age-metallicity degeneracy \citep[see, e.g.,][]{Worthey1994,Bridzius2008}. In panel b) and c), the selection of grid model metallicity shows also significant effects on the derived mass and extinction.

Comparing Fig.\,\ref{fig:1Z_vs_1Z_results_WFC3} to the pure optical case ($UBVRI$) or optical $+$ NIR case ($UBVRIJHK$) shown in Paper II, the presence of the UV filter (here UVIS1/F275W) significantly helps narrow the scatter in derived age, mass, and extinction for main diagonal panels (where the metallicity of the used grid is the same as the true metallicity) as well as for out-of-diagonal panels (where the metallicities are different). In the latter case, the similar metallicity effects are seen as in Paper II, producing offsets on the age and mass, and increasing the scatter in the extinction.

\subsection{1--metallicity vs 13--metallicities: taking metallicity effects into account}
In this test we derive age, mass, extinction, and metallicity of the same artificial clusters of fixed metallicity, but here we use a model grid containing 13 metallicities (from $\mathrm{[M/H]}$ = $+0.2$ to $-2.0$) to check the ability of the method to constrain the metallicity. We studied the clusters allowing three different amounts of photometric errors to study their influence on the derivation of parameters: Gaussian photometric errors of $\sigma$ = 0.03, 0.05, and 0.1 mag were added to each magnitude.  

	Fig.\,\ref{fig:1Z_vs_3Z_results_WFC3} shows the results; the panel b) block can be compared to the results shown in Fig.\,6 of Paper II, as the same amount of photometric error has been added to artificial cluster photometry. The presence of the ultraviolet F275W filter of the WFC3+ACS photometric system is important as it helps to significantly reduce the biases in parameter derivation, including metallicity, compared to the $UBVRI$ photometric system. 
	
Concerning photometric errors, in Fig.\,\ref{fig:1Z_vs_3Z_results_WFC3} we see a much better agreement between derived and true parameters for low and medium added photometric errors (panel a block : 0.03 mag of photometric errors, panel b block : 0.05 mag) than in the case of higher photometric errors (panel c block : 0.1 mag). 

In Fig.\,\ref{fig:1Z_vs_3Z_results_WFC3}, for each parameter, we define regions around the true value to quantify the accuracy of the method of parameter derivation for the different cases of photometric errors tested. These regions, enclosed between dashed lines in the figure, are considered regions of correct parameter derivation for this quantification of method accuracy. For the age, the region is defined up to 0.5 dex around the true value (see the dashed lines), for the mass up to 0.2 dex, for the extinction up to 0.1 mag, and for the metallicity up to 0.5 dex. In each panel, we indicate the percentage of the 10\,000 artificial clusters that have derived parameters in the correct regions. For example, in the case of photometric errors = 0.03 mag, 80\% of the clusters with true solar metallicity have metallicity derived in the correct region; the percentages are 60\% for clusters with true $\mathrm{[M/H]} = -0.6$ and 75\% for clusters with true $\mathrm{[M/H]} = -2$ (see bottom row of panel a block). When we increase the photometric errors to 0.05 mag (panel b block) or 0.1 mag (panel c block), these likelihoods of correctly deriving parameters decrease strongly, especially for intermediate metallicity. Indeed for the case of clusters with true $\mathrm{[M/H]} = -0.6$, nearly half clusters have metallicities correctly derived when photometric errors = 0.05 mag, and only 30\% when photometric errors = 0.1 mag. These numbers are only likelihoods  marginalized for each parameter separately. 

We can also compute the joint age-metallicity likelihood, i.e., the likelihood of correctly deriving the age and metallicity parameters simultaneously. For photometric errors = 0.03 mag, this likelihood is 80\%, 60\%, 75\% for clusters with true $\mathrm{[M/H]} = 0$, $-0.6$, $-2$, respectively. In this case of very high photometric accuracy, the age-metallicity likelihood is equal to the metallicity marginal likelihood (indicated in bottom of the panels block in Fig. \ref{fig:1Z_vs_3Z_results_WFC3}a) because the age derivation is very accurate, and the limiting factor is thus metallicity. For photometric errors = 0.05 mag, this likelihood is 75\%, 45\%, 65\% for clusters with true $\mathrm{[M/H]} = 0$, $-0.6$, $-2$, respectively. For photometric errors = 0.1 mag, this likelihood falls to 65\%, 25\%, 50\% for clusters with true $\mathrm{[M/H]} = 0$, $-0.6$, $-2$, respectively.

Hence, the better the photometric accuracy is, the higher the accuracy of the derived parameters can be achieved with the WFC3+ACS system filters, especially for the metallicity, as will be emphasized in the following section with real data.

\section{Application to the M31 PHAT star clusters}
\label{sec:PHAT_clusters}

\subsection{Cluster sample}

\begin{figure*}
\centering
\includegraphics[scale=0.5]{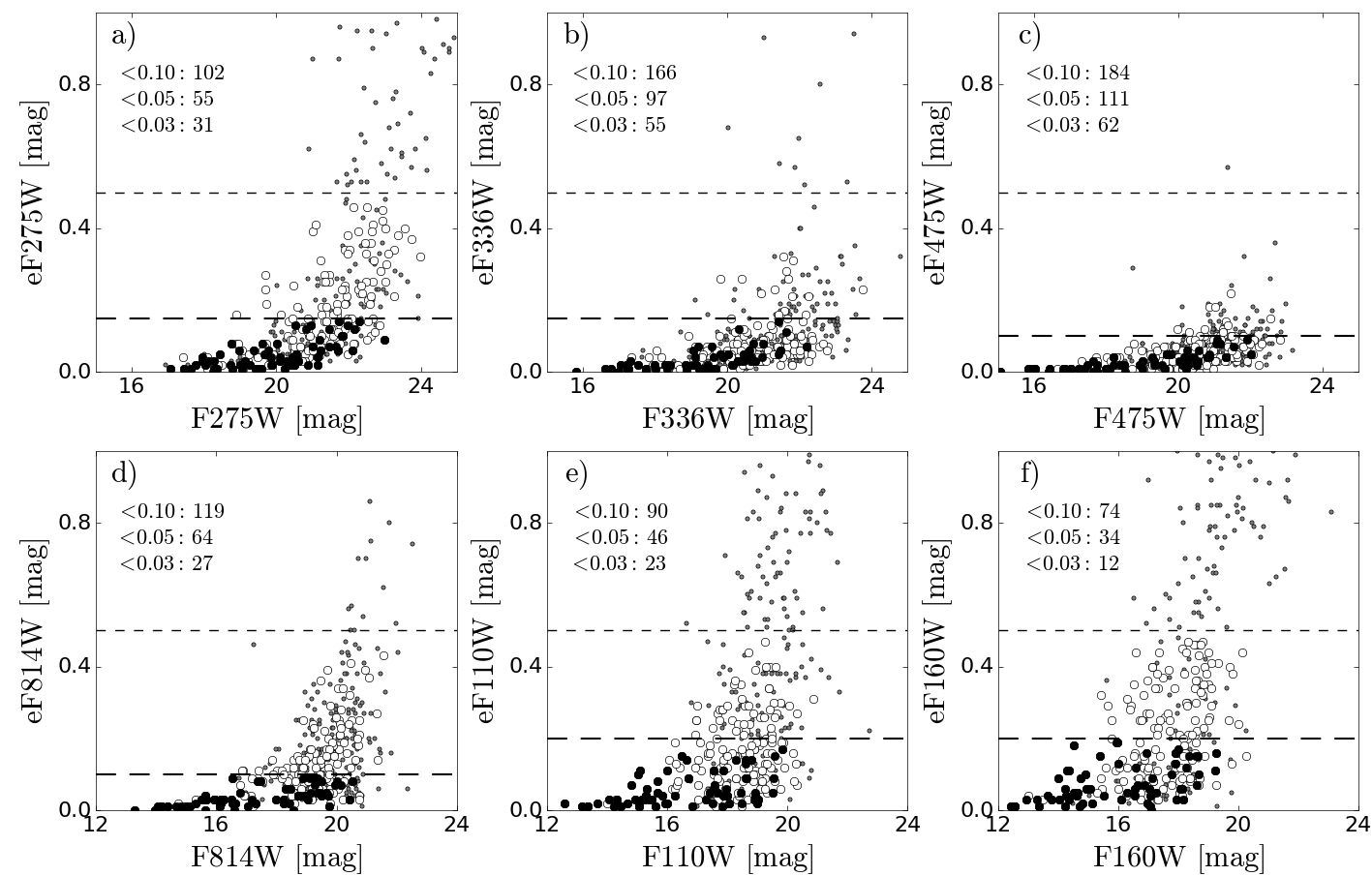}
\caption{\small Photometric accuracy of the sample of 402 clusters (all dots) selected with available photometry in all filters from the \cite{Johnson2012} catalog. Group 1 clusters (65 objects, large filled dots) have upper limit of the accuracy indicated by long dashed lines in different filters. Group 2 clusters (138 objects, large open dots) have accuracy <\,0.5 mag (short dashed lines) in all filters. Small gray dots designate clusters which are not members of group 1 or group 2 and not studied in the following because of too low photometric accuracy (they have accuracy >\,0.5 mag in at least one filter). In each panel we indicate the number of clusters (from the total 203 clusters of group 1 and group 2) with photometric accuracy better than 0.1, 0.05 and 0.03 mag in the filter associated with the panel.}
\label{fig:Photometric_errors}
\end{figure*}

Using the WFC3+ACS photometric system on board HST, the Panchromatic Hubble Andromeda Treasury (PHAT) team \cite[see, e.g.,][]{Dalcanton2012,Beerman2012,Weisz2013} performed a survey of 1/3 of the M31 galaxy, providing a large catalog of 601 clusters \citep{Johnson2012}. This catalog of star clusters was already analyzed by \cite{Fouesneau2014}, who derived their age, mass, and extinction. They used a constant solar metallicity through the whole M31 disk, arguing that the HII zone study \citep{Zurita2012} does not show any significant metallicity gradient. They allowed four different metallicities ($\mathrm{[M/H]} = -0.7$, $-0.4$, $0.0$, and $+0.4$, i.e., from Small Magellanic Cloud to super-solar metallicities) for 30 massive globular-like clusters with mass $>10^{5} M_{\odot}$.

We first selected a sample of 402 clusters from the catalog of \cite{Johnson2012} with available magnitudes in all photometric filters. Then from this sample we selected two cluster groups, for which we display the photometric accuracy for each filter in Fig.\,\ref{fig:Photometric_errors}. In cluster group 1 (65 objects, the large filled dots in Fig.\,\ref{fig:Photometric_errors}), the photometric accuracy of objects is <\,0.15 mag in F275W and F336W, <\,0.1 mag in F475W and F814W, and <\,0.2 mag in F110W and F160W. In cluster group 2 (138 objects, the large open dots in Fig.\,\ref{fig:Photometric_errors}), the photometric accuracy in each filter is <\,0.5 mag. In total, we analyze 203 clusters. In Fig.\,\ref{fig:Photometric_errors} we indicate for each filter the number of objects from the 203 clusters for which photometry is more accurate than 0.03, 0.05, 0.1 mag.

We derived the parameters of both cluster groups 1 and 2 by firstly fixing the metallicity of all clusters to the solar value, and a second time by allowing a large range of metallicities in the model grid with 13 values of $\mathrm{[M/H]} = +0.2$ to $-2.2$, in steps of 0.2. 

\begin{figure*}
\centering
\includegraphics[scale=0.5]{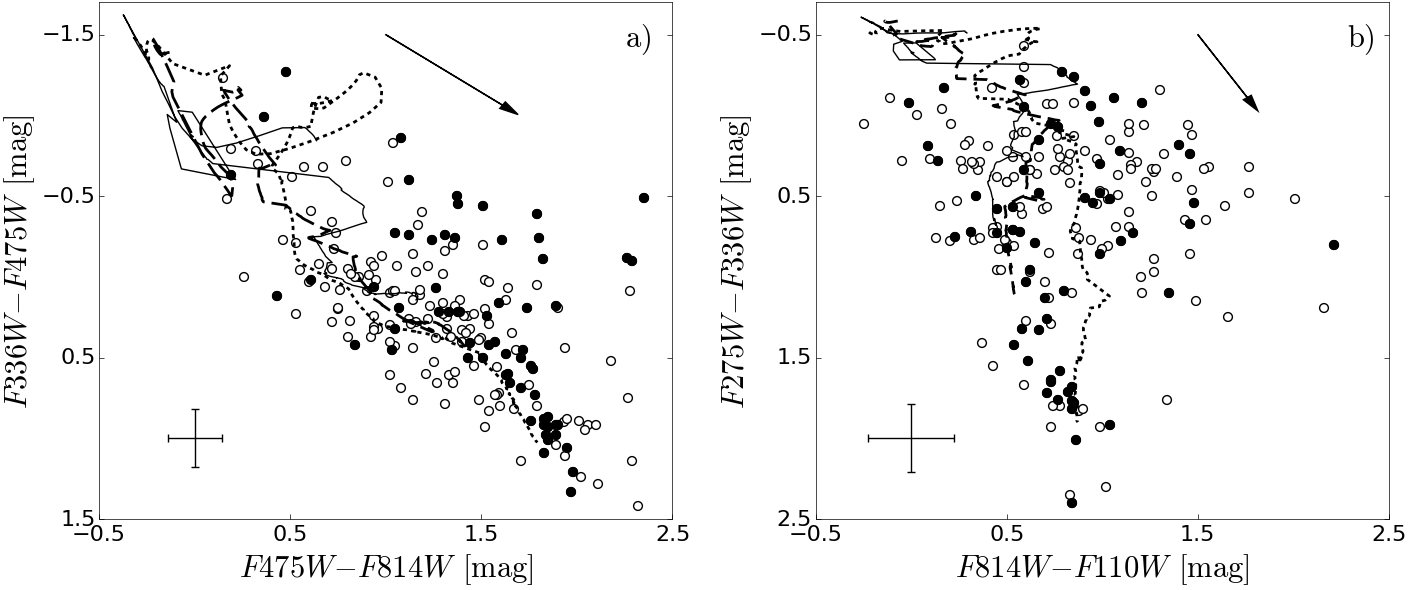}
\caption{\small Color-color diagrams of the 65 group 1 clusters (filled dots) and 138 group 2 clusters (open dots) defined in Fig.\,\ref{fig:Photometric_errors}. Panel \textbf{a)} shows the situation in optical colors only, while panel \textbf{b)} shows the situation when we make use of the ultraviolet and near-infrared filters. In both panels, the error bars indicate the maximum photometric error for the group 1 clusters. The three lines show SSP evolutionary tracks of metallicity $\mathrm{[M/H]} = 0$ (dotted line), $-1$ (dashed line), and $-2$ (solid line). The SSP ages extend from $\log_{10}(t/\mathrm{yr})=6.6$, in the upper part of each panel, to 10.1 in the lower part of each panel. The arrows indicate the direction of the extinction $A_{V}=1$.
}
\label{fig:ColourColourDiagrams}
\end{figure*}

The color-color diagrams of cluster groups 1 and 2 are shown in Fig.\,\ref{fig:ColourColourDiagrams} in optical-only filters (panel a), and UV-optical-IR filters (panel b), along with the SSP evolutionary tracks (tracing the age of clusters, valid for massive clusters only) of three metallicities as an illustration. In this figure, one can guess the advantage of using UV and IR photometry to derive the cluster metallicity. In panel a), with optical-only filters, we see that the SSPs are rather close, while this is not the case anymore in panel b), where UV and IR filters are shown. This indicates that, at least for massive and old clusters (center to bottom of panel b), the derivation of metallicity is possible with the WFC3+ACS system, provided that the photometric  accuracy is reasonable, such as for group 1 clusters (for which the maximum photometric accuracy is indicated in the Fig.\,\ref{fig:ColourColourDiagrams}). 

\subsection{Results with fixed solar metallicity}
\begin{figure*}
\centering
\includegraphics[scale=0.5]{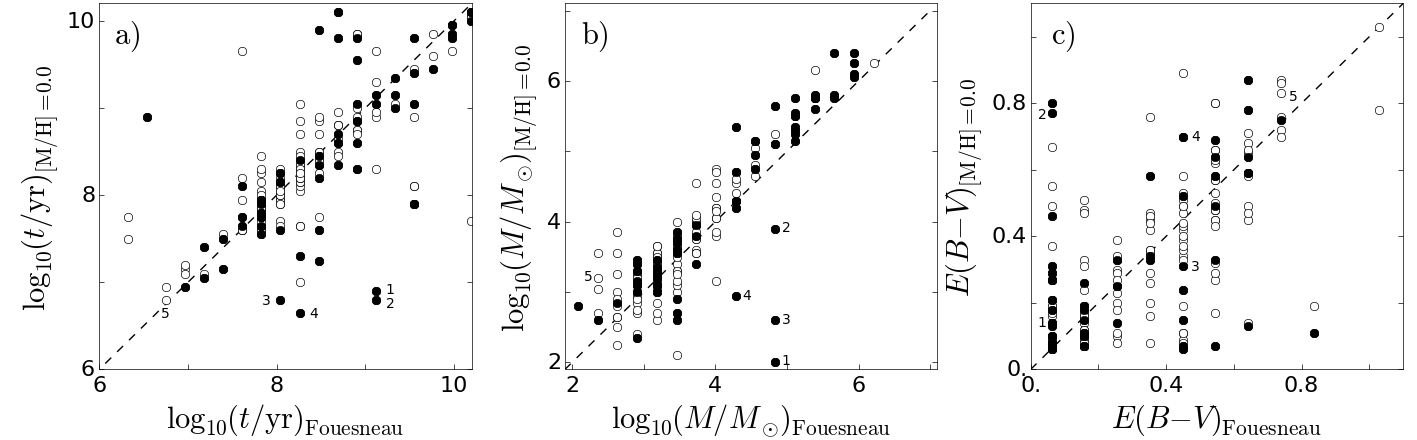}
\caption{Age, mass, and extinction derived with fixed solar metallicity grid vs the results of \cite{Fouesneau2014}. Filled dots are the 65 clusters of group 1 and open dots are the 138 clusters of group 2, which are specified in Fig.\,\ref{fig:Photometric_errors}. Clusters marked 1--5 in this figure and Figs. \ref{fig:results_free_Z_vs_fixed_Z} and \ref{fig:comparison_with_Caldwell} are the same objects.}
\label{fig:results_fixed_Z}
\end{figure*}

The age, mass, and extinction derived with the fixed metallicity grid $\mathrm{[M/H]}$ = 0 are compared to the results of \cite{Fouesneau2014} in the first row of Fig.\,\ref{fig:results_fixed_Z}. A general satisfactory agreement is observed for the majority of clusters between the parameters derived in this study and those supplied by \cite{Fouesneau2014}, except for a few objects. There are two reasons for the deviating clusters: (1) for 30 old and massive clusters, \cite{Fouesneau2014} derived parameters leaving the metallicity free to vary in the range $\mathrm{[M/H]}$= [$-0.7$,$+0.4$], while we used fixed solar metallicity, and (2) the stellar models used to derive cluster parameters are very different. 
Indeed, PEGASE-based\footnote{http://www2.iap.fr/pegase/} stellar models \citep{Fioc1999} are based on the \cite{Bertelli1994} stellar models with a simplified analytic description of \cite{Groenewegen1993} for the TP-AGB phase while PADOVA stellar models used here attempt to numerically reproduce the physics of that stellar phase \citep[see][]{Marigo2008,Girardi2010}.
For the masses, \cite{Fouesneau2014} provide the present-day mass, while the masses output by our method are initial masses, causing a natural slight shift between their mass predictions and our mass predictions. The four clusters of group 1 (Fig.\,\ref{fig:results_fixed_Z}b; filled points) strongly below the identity line and classified as massive by \cite{Fouesneau2014} are in fact globular-like clusters\footnote{Verified inspecting the images provided by PHAT: http://archive.stsci.edu/pub/hlsp/phat/}, wrongly classified by our method when the metallicity is fixed to the solar value in the model grid. We provide more details on them in the following.

\begin{figure*}
\centering
\includegraphics[scale=0.5]{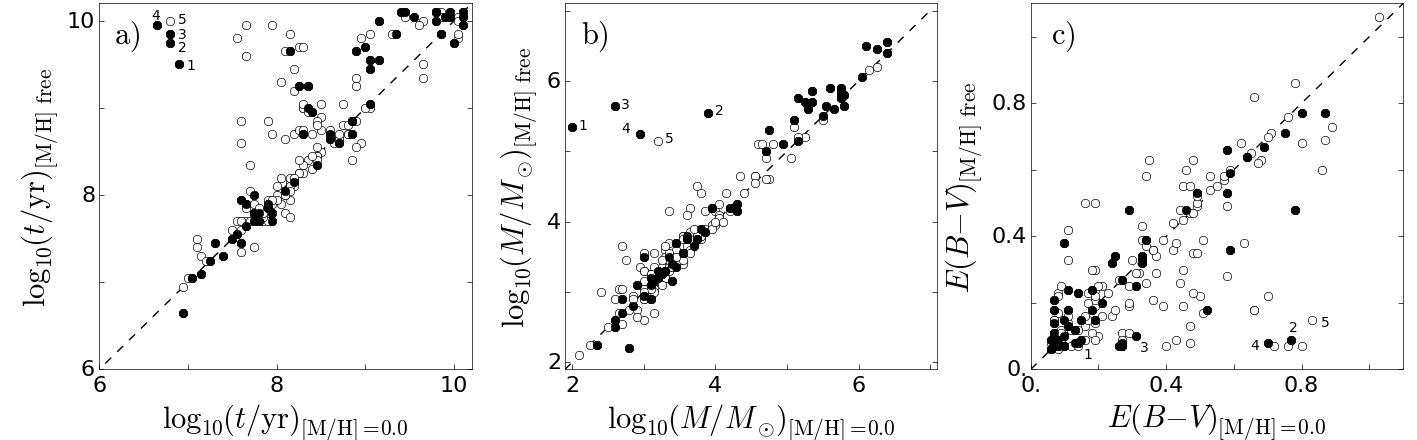}
\caption{Age, mass, and extinction  using free metallicity grid vs the results derived using a model grid of fixed solar metallicity. Filled dots are the 65 clusters of group 1 and open dots are the 138 clusters of group 2, which are specified in Fig.\,\ref{fig:Photometric_errors}. Clusters marked 1--5 in this figure and Figs. \ref{fig:results_fixed_Z} and \ref{fig:comparison_with_Caldwell} are the same objects.}
\label{fig:results_free_Z_vs_fixed_Z}
\end{figure*}

\subsection{Results with free metallicity}

Parameters derived when metallicity is left free are shown in Fig.\,\ref{fig:results_free_Z_vs_fixed_Z}. As expected from artificial tests, the introduction of the free metallicity parameter introduces a new level of complexity. Here we compare the age, mass, and extinction derived when metallicity is left free versus the same parameters when metallicity is fixed to the solar value. 

Many clusters seen as young or middle-aged in metal-fixed case are now seen as older. For example, the five points in the top left corner of the age panel (Fig.\,\ref{fig:results_free_Z_vs_fixed_Z}a, four of which are the filled points below the identity line in Fig.\,\ref{fig:results_fixed_Z}b) are seen as young in fixed solar metallicity case, but old in the free metallicity case. They are also more massive, located well above the identity line in the mass panel Fig.\,\ref{fig:results_free_Z_vs_fixed_Z}(b). These five objects are classified as low-metallicity when the metallicity is left free, which is also found likely when inspecting the images, as they are brighter in F275W and F336W filters than other globular-like clusters classified with higher metallicity. We used the individual cluster pictures in the six WFC3+ACS filters and also Sloan Digital Sky Survey (SDSS), Two Micron All Sky Survey (2MASS), and Galaxy Evolution Explorer (GALEX) images available through ALADIN\footnote{http://aladin.u-strasbg.fr/} Sky Altas to confirm that these objects are really globular-like clusters, and not young low-mass clusters. These five objects, with given PHAT ID 1439, 428, 680, 683, and 1396, are also confirmed as globular clusters in the Revised Bologna Catalog \citep[2012, version 5, see also][]{Galleti2004} with designation B064D-NB6, B229-G282, B165-G218, B167-G212, and NB21-AU5, indicated from 1 to 5 respectively in the Figs.\,\ref{fig:results_fixed_Z}, \ref{fig:results_free_Z_vs_fixed_Z}, and \ref{fig:comparison_with_Caldwell}.

\begin{figure*}
\centering
\includegraphics[scale=0.5]{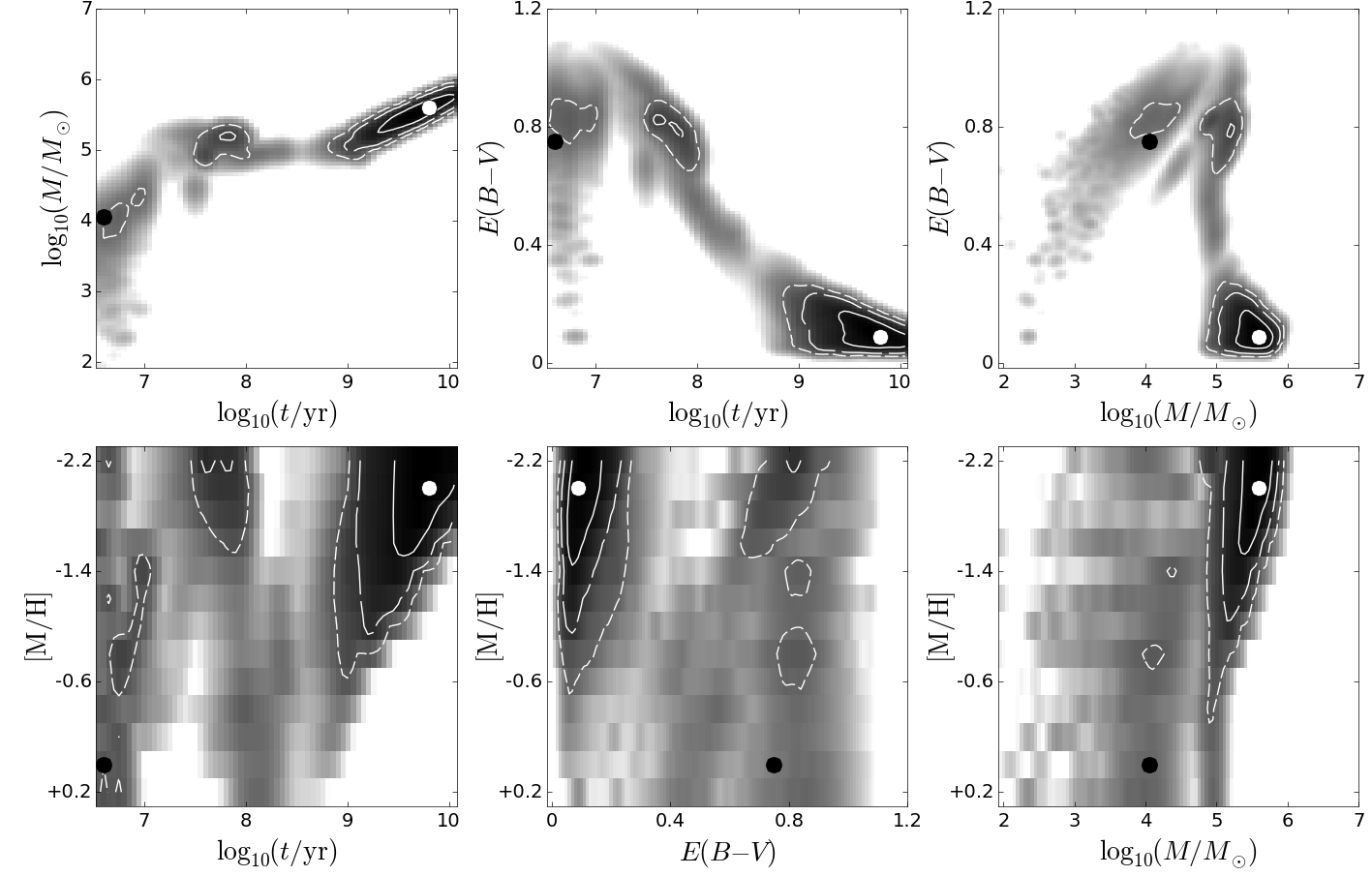}
\caption{\small Two--dimensional marginalized likelihood $L_{\mathrm{node}}$ (see Eq. \ref{eq:node_likelihood}) parameter maps derived with free metallicity model grid for the cluster ID\,428 \citep[B229-G282 in Revised Bologna Catalog V5,][indicated as ``2'' in the Figs.\,\ref{fig:results_fixed_Z}, \ref{fig:results_free_Z_vs_fixed_Z}, and \ref{fig:comparison_with_Caldwell}]{Galleti2004} photometry taken from \cite{Johnson2012} catalog. The white points indicate the maximum $L_{\mathrm{node}}$ in the 4--dimensional parameter space, while the black points show the solution when the metallicity is fixed to solar value. The white contour lines contain 68\% (solid line), 95\% (long-dashed line), and 99\% (short-dashed line) of the marginalized likelihood.}
\label{fig:ID428_proba_maps}
\end{figure*}

\begin{figure*}
\centering
\includegraphics[scale=0.495]{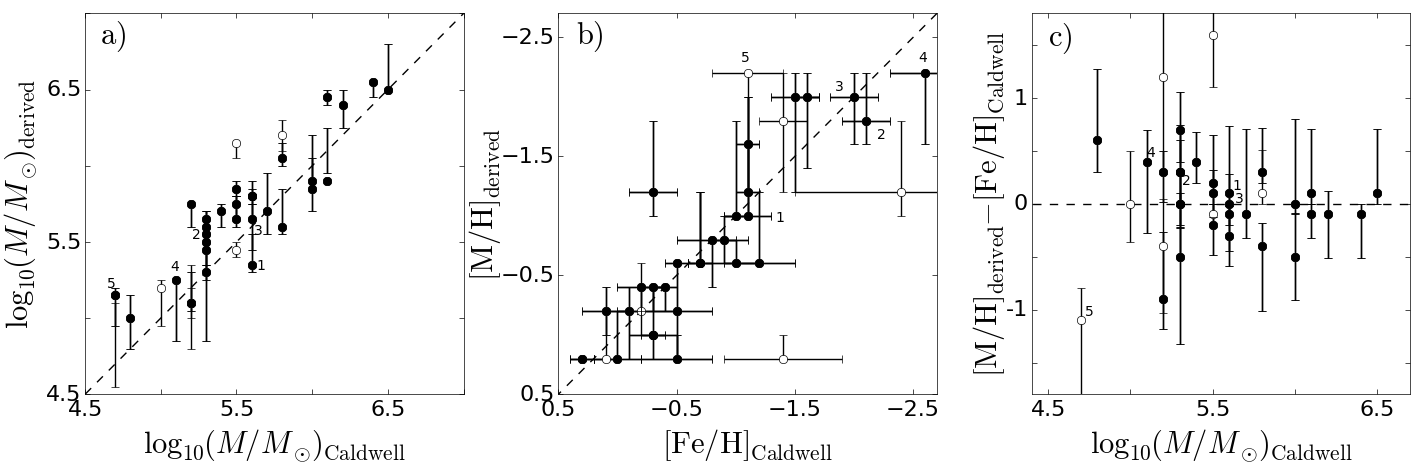}
\caption{\small Our results vs those of \cite{Caldwell2011} for the 36 common massive clusters, shown for the mass (panel \textbf{a}), the metallicity (panel \textbf{b}). Panel \textbf{c)} shows the difference of metallicity between this work and \cite{Caldwell2011} values vs the cluster mass. For the panel \textbf{c)}, error bars of \cite{Caldwell2011} and derived here are summed quadratically. Filled dots are the group 1 clusters and open dots are the group 2 clusters, as defined in Fig.\,\ref{fig:Photometric_errors}. Clusters marked 1--5 in this figure and Figs. \ref{fig:results_fixed_Z} and \ref{fig:results_free_Z_vs_fixed_Z} are the same objects.}
\label{fig:comparison_with_Caldwell}
\end{figure*}

As an illustration, 2--dimensional marginalized likelihood maps are shown for one of these objects, ID 428 (indicated as ``2'' in the Figs.\,\ref{fig:results_fixed_Z}, \ref{fig:results_free_Z_vs_fixed_Z}, and \ref{fig:comparison_with_Caldwell}), in Fig.\,\ref{fig:ID428_proba_maps}. The likelihood maps are given for the cluster classification using all 13 metallicities of the model grid. The parameters derived taking the maximum of likelihood $L_{\mathrm{node}}$ (see Eq. \ref{eq:node_likelihood}) in the 4--dimensional model grid are indicated with the white points in the 2--dimensional marginalized likelihood maps. Additionally, the black dots indicate the parameters obtained when the metallicity is fixed to the solar value in the model grid (3--dimensional grid), resulting in a wrongly classified younger, low-mass, and much more extincted solution. 

We note that in Fig. \ref{fig:results_free_Z_vs_fixed_Z}(a) dozen of the objects with ages, derived assuming fixed solar metallicity, around $\log_{10}(t/\mathrm{yr})=8$ have overestimated ages when derived with free metallicities. This is very likely because these objects are relatively faint, with poor photometric accuracy, and are contaminated by bright red background stars. However, careful analysis of the WFC3+ACS object images and their likelihood maps (similar to those shown in Fig. \ref{fig:ID428_proba_maps}) allows us to resolve degeneracies in most of the cases.

Recently, \cite{Caldwell2011} produced the spectroscopic study of old star clusters of M31 galaxy, using Lick indices to derive their age, mass, extinction, and metallicity. To check the reliability of our derived metallicity, we compare those of the 36 clusters common between the \cite{Caldwell2011} sample and the clusters analyzed in this study in Fig.\,\ref{fig:comparison_with_Caldwell}. As \cite{Caldwell2011} fixed the age of most clusters to 14 Gyr, here we only compare the mass and metallicity of the clusters. Again, an overall agreement is found between the parameters. The accuracy of our photometric metallicity derivation is linked to the mass, and thus very likely to the signal-to-noise of available photometry for each object, as the scatter seen in Fig.\,\ref{fig:comparison_with_Caldwell}(c) is increasing with decreasing star cluster mass. The five clusters studied above are also indicated in Fig.\,\ref{fig:comparison_with_Caldwell}, where one can see that the metallicity derived using our method coincides well with the spectroscopic method of \cite{Caldwell2011}.

\begin{figure}
\centering
\includegraphics[scale=0.495]{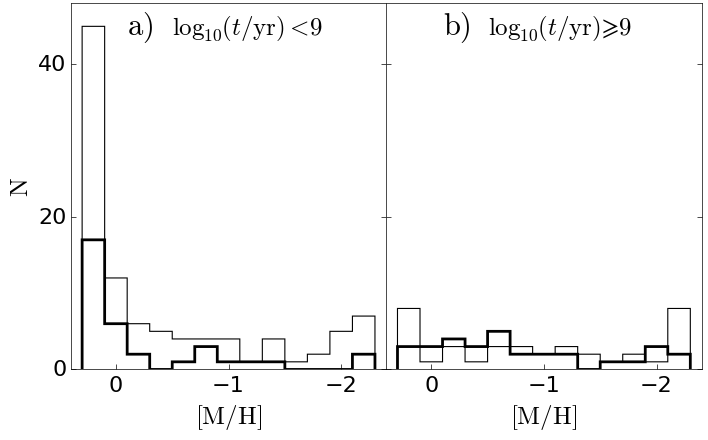}
\caption{\small The derived metallicity of clusters with ages lower than 1 Gyr (panel a), and of clusters with ages higher than 1 Gyr (panel b) for group 1 (thick line) and group 2 (thin line) clusters.}
\label{fig:metallicity_histograms}
\end{figure}

In Fig. \ref{fig:metallicity_histograms}, we show the histograms of derived metallicity of group 1 (thick line) and group 2 (thin line) clusters for young age clusters ($\log_{10}(t/\mathrm{yr})<9$, panel a) and old age clusters
($\log_{10}(t/\mathrm{yr})\geqslant 9$, panel b). Most of the young clusters are  classified as metal-rich, while the old cluster metallicities are more dispersed. Note, that even for low photometric accuracy, Fig. 9(a) still tells us that the clusters are more likely metal-rich than metal-poor. Also, it is interesting to compare Fig. \ref{fig:metallicity_histograms}(a) to the artificial tests of Fig. \ref{fig:1Z_vs_3Z_results_WFC3}, in the worst photometric accuracy (panel c block). Indeed, we see a strong similarity between the case with true $\mathrm{[M/H]} = 0$ and the left panel of Fig. \ref{fig:metallicity_histograms}. Therefore, fixing the metallicity to a value as high as the solar value assumed by \cite{Fouesneau2014} is likely a good choice for most of the young M31 clusters. Note that \cite{Caldwell2009} recommends supersolar metallicities for young M31 clusters, although they did not derive individual young cluster metallicities.

\section{Conclusions} 
\label{sec:conclusions}

We demonstrated, with the use of artificial star clusters, that the derivation of the age, mass, and extinction is possible when using the WFC3+ACS photometric system, allowing us to avoid the biases introduced when the metallicity is fixed. In Paper II, we showed that optical (UBVRI) or optical+NIR (UBVRIJHK) photometry cannot achieve the parameter derivation accuracy reached by using the WFC3+ACS photometric system; the improvement in the latter system is because of the presence of the UV filter UVIS1/F275W of WFC3. In the case when the photometric errors are reduced up to 0.05 mag for each filter, an accurate derivation of the metallicity parameter is also possible. 

Using a sample from the M31 PHAT ``year 1'' star cluster catalog of \cite{Johnson2012}, we derived their parameters using fixed solar metallicity consistent with the \cite{Fouesneau2014} study, and also with free metallicity in the model grid. For the clusters in common with the \cite{Caldwell2011} spectroscopic study, we compare the metallicity with their values and found and overall agreement. This demonstrates, with the use of real star clusters, that the WFC3+ACS photometric system is fit to evaluate the star cluster parameters when the metallicity is unknown, and evaluate the metallicity when the signal-to-noise of photometry is high enough.

\begin{acknowledgements}
We are grateful to the anonymous referee who helped improve the paper. This research was funded by a grant (No. MIP-074/2013) from the Research Council of Lithuania.
\end{acknowledgements}

\bibliographystyle{aa}

\end{document}